\documentclass[a4paper,fleqn]{cas-dc}
\usepackage[numbers]{natbib}
\usepackage{subcaption}
\usepackage{graphicx}
\usepackage{color,soul}

\def\tsc#1{\csdef{#1}{\textsc{\lowercase{#1}}\xspace}}
\tsc{WGM}
\tsc{QE}
\tsc{EP}
\tsc{PMS}
\tsc{BEC}
\tsc{DE}

\begin{document}
\let\WriteBookmarks\relax
\def\floatpagepagefraction{1}
\def\textpagefraction{.001}

\shorttitle{ }

\shortauthors{Mirmojtaba Gharibi and John-Paul Clarke}

\title [mode = title]{On the number of freeway lanes and its positive or negative effect on safety}                      
\tnotemark[1]

\tnotetext[1]{This document is the results of the research
   project funded by the National Science Foundation.}


%
\author[1]{Mirmojtaba Gharibi}[
                        orcid=0000-0003-4295-5051]

\cormark[1]

\fnmark[1]

\ead{mgharibi@uwaterloo.ca}


\credit{Conceptualization, Data curation, Formal analysis, Methodology,
Software, Roles/Writing - original draft}

\affiliation[1]{organization={The University of Texas at Austin},
    country={United States}}

\author[1]{John-Paul Clarke}[style=chinese]

\credit{Methodology, Supervision, Roles/Writing - review \& editing, Resources, Funding acquisition, Project administration}

\cortext[cor1]{Corresponding author}



\ExplSyntaxOn
\cs_gset:Npn \__first_footerline:
  { \group_begin: \small \sffamily \__short_authors: \group_end: }
\ExplSyntaxOff 

\begin{abstract}
We address the 80-year-old question of whether a freeway with more lanes results in fewer or more accidents. For finding the optimally safe number of lanes, in particular, we look at three types of accidents that are prevalent on urban freeways, namely "following too closely", "driver inattention", and "unsafe change of lanes". To do so we extend the intelligent driver model ($\text{IDM}$) to create a microscopic traffic flow model which is capable of producing accidents. We study the rate of accidents relative to a baseline 2-lane unidirectional freeway via Monte Carlo simulation. For each simulation instance we create a starting configuration involving only a few cars over a short segment of the freeway and simulate the dynamics thereafter. Furthermore, we look at the number of shoulders present, and show that the presence of shoulders can positively or negatively affect the accident rate depending on the type of accident. 
\end{abstract}



\begin{keywords}
Microscopic traffic model \sep Freeways safety \sep Accident analysis \sep Intelligent driver model (IDM) \sep Simulation
\end{keywords}

\maketitle

\section{INTRODUCTION}

As road congestion in different cities gets worse, a common solution is to build multi-lane freeways or add capacity to the existing freeways. However, the effect of such decisions on safety is under-researched \cite{ahmed2015evaluation} despite the fact that the relationship between the number of lanes on a freeway and the safety of that freeway has been the source of debate between practicing engineers/planners and the majority of researchers since the modern road-building started almost 80 years ago. While the former group believes that the reduced congestion results in safer freeways, the latter group believes that adding lanes will reduce safety.  \cite{kononov2012relationship, ahmed2015evaluation}.


The main question addressed in this paper is `How does the addition of lanes (compared to a baseline two-lane uni-directional freeway) affect the rate of different types of accidents?'; specifically "following too closely", "driver inattention", and "unsafe change of lanes," the most prevalent types of accidents according to the \cite{txdot2021crash}. 
In particular, we investigate what setting achieves the maximally safe utilization of a freeway. This knowledge can help with the development of policies that reduce specific types of accidents directly by choosing road characteristics and traffic planning that maximizes the safety.

Most accident prediction studies rely on historical crash data to predict accident rates \cite{mannering2014analytic}. There are a few major drawbacks to these models. Firstly, data collection and validation are costly. Secondly, data from specific roadways geometries, conditions, etc., may not apply to a new unique situation \cite{tan2012development}. Thirdly, the quality of the crash data that is collected is adversely affected by issues such as under-reporting of less severe accidents, low sample size, and time-varying factors to name a few \cite{lord2010statistical, zheng2021modeling}. 
Furthermore, most of the few studies using this approach, study only the effects of expansion from 2-lane to 4-lane bidirectional freeways. However, a bidirectional 2-lane freeway lacks a dedicated passing lane in each direction and is thus characteristically different from the highways with more lanes which are the focus of this paper.

 An alternative approach proposed by some researchers is the use of microscopic models. Most microscopic models for the study of safety have been developed with other scenarios in mind, such as the effect of work zones on accident rates and conflicts in the intersections. Thus, there are very few models that can be used to study the effect of the number of lanes on safety. Furthermore, many of these models do not produce accidents and instead resort to studying the rate of "near-misses" as a surrogate for the rate of accidents \cite{zheng2021modeling}. 

We address these shortcomings by extending a well-known car following (microscopic) model, the intelligent driver model ($\text{IDM}$) together with the lane changing model called MOBIL (short for minimizing overall braking induced by lane change). As alluded to above, traffic in the aforementioned models is controlled so as to avoid accidents. Thus, we had to extend them to make them suitable for studying accidents. 

We generate random traffic scenarios involving a few cars for a short duration. For these short simulations, depending on which of the named three accident types are studied, a triggering event will cause a potential conflict. We observe the number of crashes caused by these conflicts. A benefit of such a model compared to  statistical modeling is that it provides insights into the accident phenomenon which can be directly used by policymakers. 

\begin{figure*}
    \centering
    \begin{subfigure}[b]{0.95\textwidth}
        \centering
        \includegraphics[trim=0 0 0 0,clip,width=\textwidth]{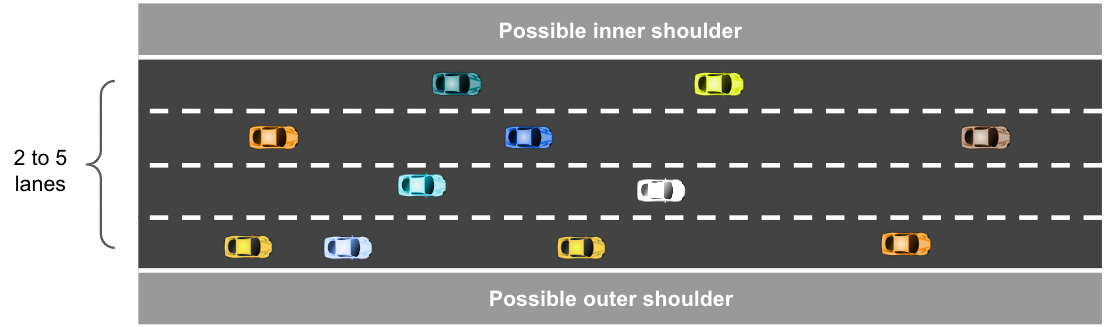}
    \end{subfigure}

    \caption[ The average and standard deviation of critical parameters ]
    {\small A microscopic model extended from $\text{IDM}$ is used to study how the relative rates of accidents (of certain types) evolve as the number of lanes increases from 2, to 3, 4, and 5 lanes in a uni-directional freeway. In doing so, we also examine all combinations of shoulders or lack thereof. } 
    \label{fig:micro}
\end{figure*}

Our contribution can be summarized as follows. We have created a microscopic model based on the $\text{IDM}$ and MOBIL models that can produce accidents and can be used to study the safety of vehicle transportation. In the context of this paper, we use this general tool to study the effect of increasing the number of lanes on safety in order to find the optimally safe number of lanes (Figure \ref{fig:micro}). We are not aware of any studies in that area, using a similar tool. Furthermore, rather than analyzing accident rates in an aggregated way as it is commonly done, we study the effect of lane expansion on three prevalent accident types in isolation.

By simulating small scenarios with only a few cars for a short duration, our methodology produces many samples with relatively low computational cost with enough of them resulting in accidents to draw statistical inference from the observed data. This allowed us to directly study the rate of accidents rather than surrogate metrics such as the number of near-misses.

Another contribution we have made is the explicit study of the effect of shoulders as we add more lanes. In most existing studies, there is no mention of how many (if any) shoulders exist on the freeway. This turns out to be an important factor since, in our study, we show that adding more lanes to a freeway with two shoulders may reduce safety for certain accident types while improving safety in a freeway with no shoulders for the same accident type. 

\subsection{Related work}

The authors of \cite{kononov2012relationship} develop a non-linear model for the relation between the flow parameters and the observed crash rate for specific segments of two 4-lane (bidirectional) freeways near Denver, Colorado. Their goal was to determine how adding lanes affected the crash rate. Based on their model, within the ranges of interest for the model parameters, if the density goes down, it will result in fewer accidents. So they conclude, that the decision to add 2 more lanes will result in safer freeways. The authors speculate that adding more lanes will result in more lane-change conflicts, but expect (without further justification) that this increase is not enough to negate their earlier conclusion. A shortcoming of this study in our opinion is the assumption that a model created for a 4-lane freeway will generalize well to a 6-lane freeway. In \cite{kononov2008relationships}, the authors use a neural network to develop safety performance functions (SPF) to predict accident rates based on the annual average daily traffic (AADT) of a few freeways with a different number of lanes. These SPFs are then used to extrapolate the accident rates by the hypothetical addition of extra lanes. Therefore, a fundamental difference and improvement in our work compared to the two cited papers is that we run our microscopic model for each number of lanes and draw conclusions about the safety based on the directly observed results. Furthermore, the 2-lane freeway exhibits a completely different dynamic in car passing compared to freeways with more lanes as is the focus of our study.

In \cite{ahmed2015evaluation}, the authors study the safety effect of converting urban and rural 2-lane roadways to 4-lane divided roadways. They find a substantial drop in the accident rate. However, it is worthy to mention that the AADT does not show a significant change (less than 14\%) during the period the authors study the treated sites. This is perhaps one of the first studies to take a before-after approach. They also combine this with a comparison with a group of untreated sites to account for factors such as long-term traffic trends. To drive their result, they compare 4 methods; namely before-after studies with and without using the comparison groups as well as empirical Bayes (EB) and Bayesian approaches. A similar study was done in \cite{asare2014safety} for 2-lane to 4-lane conversions in Wisconsin using simple before-after (i.e. with no comparison group) and EB showing a 7\% to 82\% reduction in accidents. 

In \cite{council1999safety}, citing a lack of before and after data on how the conversion from 2 lanes to 4 lanes affects the safety of rural roads, the authors develop a model using cross-sectional analysis of data for existing 2-lane and 4-lane roads. 

In \cite{daniel2011relating} the authors estimate the accident rates based on the roadway capacity using both Poisson and binomial regression models developed. To estimate the capacity of a roadway (in which the number of lanes is a major factor), the authors use the framework laid out in Highway Capacity Manual (HCM) \cite{manual2010hcm2010}. They conclude that higher capacity results in higher accident rates.

Some studies look at the number of lanes among many other factors to estimate the accident rates. In \cite{chen2021analysing}, the authors study the relationship between commercial vehicle mix and safety. In the Bayesian random-parameter Tobit model developed, the authors also consider secondary factors such as the highway geometric properties and conclude that a higher number of lanes among other things is associated with higher crash rates for commercial vehicles. In \cite{ahmed2011exploring}, to predict crash rates in a mountainous roadway, 
the authors develop a Bayesian hierarchical model that takes into account various variables including the number of lanes. In particular, they show for the studied roadway, a 40\% reduction in accidents is predicted in segments with three lanes compared to the segments with two lanes. In \cite{park2010evaluating}, the authors develop a negative binomial regression model to predict accident rates based on mainly ramp density and horizontal curves. The parameters for the statistical model parameters also included the number of lanes. However, the separate models that were created for each lane category produced inconsistent results concerning the safety as a function of ramp density and horizontal curves (for 6-lane, the accident rate decreased, then increased again for 8-lane). In \cite{gaweesh2019developing} the authors develop crash prediction models for rural mountainous roadways based on three different methods, namely, the negative binomial (BN) regression model, partial autoregressive (SAR) model, and non-parametric multivariate adaptive regression splines (MARS) model. As one of the variables, the number of lanes is also included, but its effect is not discussed.


In the study of traffic safety, it is natural to attempt to utilize the large body of work on microscopic traffic flow models. However, these models are designed for normal driving behavior and will not result in collisions. As a workaround, the researchers have looked at various metrics that represent a risky interaction such as time to collision (TTC), post encroachment time (PET), maximum deceleration rate (DR), and maximum speed as surrogates for safety performance to enable them to make predictions about the number of accidents in the real world \cite{duong2010calibration, cunto2008calibration, gettman2003surrogate}. Similar to our work, the authors in \cite{gettman2003surrogate} consider various conflict events (albeit for intersections rather than freeways) and collect data regarding surrogate safety performance indicators from microscopic simulations which are different from our work where we deal with the collision data directly.

There is no shortage of car-following models in the literature. Generally, the norm is to set the parameters so that no accidents will occur. One such classic model is the intelligent driver model (IDM)  \cite{treiber2000congested, kesting2010enhanced}. The non-collision principle poses a challenge when using such models to study accident rates. By adjusting the parameters, it is possible to have models that perform well for the most part but occasionally result in accidents. The authors in \cite{treiber2006delays} develop a meta-framework for adding other typical human behavior factors that are missing from some classical models to these models (e.g. reaction delay, perception errors, etc.). It is shown that if reaction times are too long, the model will produce crashes.

In \cite{yue2020low} the authors generate accident risk scenarios such as front collision, rear-end collision, etc., using a microscopic model as a proxy for costly and time-consuming real data collection.

In \cite{wang2018combined} the authors use a microscopic model together with the extreme value method (EVT) to predict crash rates from the frequency of near misses according to the safety performance surrogate metrics for ten urban intersections in China. The EVT method is used widely in finance, insurance, and risk management to approximate the probability of rare events.





In \cite{hou2020study} the authors use a cellular automata-based microscopic model to predict the single-vehicle and multi-vehicle crash rates in work zones during extreme weather. In the cellular automata model, a reaction time of $1$s is considered for all drivers. In the case of multi-vehicle crashes, two types of crashes are considered. The first accident type is triggered by a sudden deceleration by a leading vehicle. The second type of accident investigated is merging onto the same target lane by two vehicles which can result in a rear-end collision. 

In \cite{marzoug2017car} the authors estimate the accident rate at a junction by using the Nagel-Schreckenberg model (NS) \cite{nagel1992cellular} which is a probabilistic cellular automaton model. They factor in the probability that the two drivers approaching the junction either cooperate or do not cooperate. 

In \cite{pang2015simulation} the authors extend an NS model to three lanes and include effects such as low visibility and driver's reaction time. The combined effect of this is a model that unlike NS can result in accidents. The authors study the fraction of vehicles involved in an accident as the visibility decreases. 

In \cite{boccara1997car} the authors pioneer to study the effect of drivers not respecting the safety distance rule in the NS model in causing accidents. In \cite{mhirech2015effect}, using the same model, the authors study the effect of a traffic light in the middle of the road and its cycle time on the number of accidents. In \cite{moussa2003car} the authors extend the work of \cite{boccara1997car} to study two scenarios where there is a sudden deceleration by the car ahead or a sudden stop. The authors study the accident rate as the density increases for the one-lane roadway. In \cite{moussa2005dangerous}, the same author studies the probability of accidents using a similar NS model for two lanes. For the lane change mechanism, the authors use the rules developed in \cite{nagel1998twolane, knospe1999disorder, knospe2002realistic}.

The authors of \cite{paschalidis2019combining}, extend the GM car following model \cite{gazis1961nonlinear} to study the effect of stress on the drivers' behaviors. They show the acceleration rates will be higher for the drivers with stress, akin to the behavior of the so-called aggressive drivers in the model. The study can indirectly shed light on traffic safety and potential accident rates by studying arguably unsafe driving behaviors.


\section{Method}
\subsection{Base microscopic model}
As stated previously, we use a classic car following (microscopic) model called the  intelligent driver model ($\text{IDM}$) as our starting point \cite{kesting2010enhanced, treiber2000congested}. We use MOBIL lane-changing model \cite{kesting2007general} to extend the $\text{IDM}$ to a multiple lane model. 

In the $\text{IDM}$ model the current acceleration is given by 

\begin{equation}
\label{eqBaseIDM+}
    a(s,v,\Delta v)=A\left[1-\left(\dfrac{v}{V_0}\right)^\delta-
    \left(\dfrac{s^*(v,\Delta v)}{s}\right)^2\right].
\end{equation}
The variables in Eq. \ref{eqBaseIDM+} are
\begin{itemize}
    \item gap with the vehicle at the front: $s$
    \item current velocity: $v$
    \item approach rate to the vehicle at the front: $\Delta v$
    \item the minimum desired gap: $s^*$ 
\end{itemize}
where $s^*$ is calculated by 
\begin{equation}
    s^*(v,\Delta v)=S_0+vT+\dfrac{v\Delta v}{2\sqrt{AB}}.
\end{equation}

Furthermore, the parameters of the model are defined and set according to \cite{kesting2010enhanced} as follows. 

\begin{itemize}
    \item desired speed $V_0$: $120km h^{-1}$
    \item free acceleration exponent $\delta$: $4$
    \item desired time gap $T$: $1.5s$
    \item jam distance $S_0$: $2.0m$
    \item maximum acceleration $A$: $1.4ms^{-2}$
    \item desired deceleration $B$: $2.0ms^{-2}$
\end{itemize}

To transform the $\text{IDM}$ to a multi-lane model, we used a simplified version of the MOBIL model where each driver is purely selfish, maximizing only their  gain. Since we are simulating the model for only a short duration, it is plausible lane changing will be used mainly to evade accidents and hence this appears to be a justified simplification. 

The rule for changing the lane for a vehicle is as follows. Firstly any lane change must be safe for the assumed follower vehicle in the new lane. Secondly, it must be advantageous to switch lanes. More formally, these two conditions are formulated as 
\begin{equation}
\label{eqSafeLaneChange}
    a_\text{follower} \geq -B_\text{safe}
\end{equation}
and
\begin{equation}
    a_\text{new} - a_\text{cur} > A_\text{thr}
\end{equation}
where variables in the equations are
\begin{itemize}
    \item hypothetical acceleration of the follower in the new lane: $a_\text{follower}$
    \item new acceleration after switching lanes: $a_\text{new}$
    \item current acceleration: $a_\text{cur}$.
\end{itemize}
Furthermore, we set and define the parameters above as follows.
\begin{itemize}
    \item maximum safe braking deceleration  $B_\text{safe}$: $4ms^{-2}$ (similar to the original MOBIL model \cite{kesting2007general})
    \item acceleration gain threshold  $A_\text{thr}$:  $0.2ms^{-2}$
\end{itemize}

\subsection{Extension to the basic model}


We extend the base microscopic model to make it suitable for studying accidents via the following additions and modifications. 

\begin{itemize}
    \item risk factor parameter: $R$ with $R\in[0,1)$ 
    \item lane number: $l$ is a real number with $l\in[1,N]$ where $N$ is the number of lanes (one-way).
    \item lane change progress: $p$ with $p\in[0,1)$
    \item state indicator (depending on how harsh the driver needs to brake): \[\text{state}\in \{\text{safe}, \text{hazardous},\text{critical}\}.\] Each state is quantified below, merely according to the needed braking severity.  
    \item max deceleration parameter: $B_\text{state}$. The set values are: 
        \begin{equation}
    \left\{
        \begin{array}{lr}
            B_\text{safe}=4ms^{-2} \text{ (similar to the base model)},\\
            B_\text{hazardous}=7ms^{-2} \text{, and}\\
            B_\text{critical}=9ms^{-2}\text{ (theoretical maximum 
\cite{kesting2007general})}.
        \end{array}\right.
    \end{equation}
    where $B_\text{safe}$ and $B_\text{critical}$ are set according to the literature and $B_\text{hazardous}$ is set as an intermediate value between the two extremes.
    
    \item state indicator: state. For a vehicle with real valued acceleration ($a\in \mathbb{R}$), at any moment, the vehicle is in one of the three states as follows depending on the braking rate of the driver (we take take the braking rate as a proxy for how critical the driver perceives the situation).
    \begin{equation}
    \text{state} = 
    \left\{
        \begin{array}{lr}
            \text{safe}, & \text{if } B_\text{safe}\leq a \leq A\\
            \text{hazardous}, & \text{if } B_\text{hazardous}\leq a<B_\text{safe}\\
            \text{critical}, & \text{if } B_\text{critical} \leq a < B_\text{hazardous}
        \end{array}\right.
    \end{equation}
    \item lane change speed parameter: $W_\text{state}$. While a realistic lane change takes between 2.4s to 5.8s according to \cite{lee2004comprehensive,hanowski2000field,tijerina1999individual, toledo2007modeling}, following the simplifying convention of many car following models with lane change (e.g., \cite{kesting2007general, wagner1997realistic}), the lane change in our model happens almost instantly as follows.
    \begin{equation}
    \left\{
        \begin{array}{lr}
            W_\text{safe}=W_\text{hazardous}=0.5\text{ lane}/s \text{, and}\\
            W_\text{critical}=2\text{ lane}/s.
        \end{array}\right.
    \end{equation}

\end{itemize}
In the model, we introduce a parameter $R$ representing the risky behavior of drivers by reducing the variable $s^*$, the minimum desired gap between the cars as follows.
    \begin{equation}
    s^*(v,\Delta v)=\left(S_0+vT+\dfrac{v\Delta v}{2\sqrt{AB}}\right)\left(1-R\right).
    \end{equation}
    
Also, we generalize the safety rule of Eq. \ref{eqSafeLaneChange} and model the lane change in a continuous way as follows.
\begin{equation}
\label{eqSafeLaneChangeState}
    a_\text{follower} \geq -B_\text{state},
\end{equation}
\begin{equation}
l_\text{cur}(t) = l_{old} + (l_\text{new}-l_\text{old})p(t),
\end{equation}
and
\begin{equation}
p(t) = W_\text{state}(t-t_0)
\end{equation}
where $t_0$ is the time the lane change started. Once the lane change is complete $p(t)$ will be set to $0$ and the current lane will be updated to $l_\text{cur} = l_\text{new}$.

As a vehicle's state becomes more dangerous (which we deduce from the braking rate of the driver), we allow more extreme evasive maneuvers for the vehicle to escape a collision. In particular, the lane change speed $W_\text{state}$ as defined above, allows for a faster lane change in a critical state. Also, a car in a hazardous or critical state is allowed to switch lanes to shoulders. A vehicle driving on a shoulder has to reduce its speed with the maximum deceleration (i.e., $B_\text{critical}$) till it reaches the speed \[v_\text{shoulder}=\text{min}\left(7ms^{-1}, \text{max}\left(v_\text{avg}-1,0ms^{-1}\right)\right)\] where $v_\text{avg}$ is the average speed of all vehicles. This is to make sure no driver will take advantage of driving on the shoulders to avoid congestion. 

\subsection{Discussion and justifications for the model}

According to the data from Texas Department of Transportation, the annual number of kilometers travelled in Texas is 461 billions\cite{txdot2021crashFacts}. Traffic accident rates for the types "Followed too closely", "Driver inattention", and "Unsafe change of lanes" are 0.09, 0.36, 0.18 per million kilometers, respectively \cite{txdot2021crash}. Given the infrequency of accidents and due to computational constraints, we have exaggerated the parameters that induce risky behavior in our model to adequately record a significant number of accidents. Accordingly, the risk factor value $R$ incorporated into our model is deliberately inflated such that we can collect enough accident samples given the computational power at our disposal. Otherwise, replicating the absolute accident rates observed in reality would necessitate prohibitive computational resources. Thus, in our numerical results, we concentrate solely on comparing relative rates, rather than absolute ones.

We have incorporated into our model a concept known as a vehicle's 'safety state', which essentially encapsulates a driver's perception of their current safety condition. We included this element in line with the general understanding that drivers, when perceiving a critical situation, are likely to perform more drastic evasive actions such as rapid lane changes or abrupt braking or switching to the shoulders. The 'safety state' we have defined is solely predicated on the severity of braking required to avoid a collision, if feasible. We use the severity of the braking as a proxy for the driver's perception of their safety. This approach aligns with well-established models like \cite{kesting2010enhanced}, where the rate of braking varies based on the criticality of the situation. The inclusion of this notion in our model is intended to modulate the speed of lane changes and the potential use of the shoulders. 

Although the original IDM model is accident-free given any safe spacing between vehicles, potential for accidents still exists in our enhanced model. This is due to the cap we have set on the acceleration rate at a minimum of $-9 m/{s^2}$ (approximately the  maximum limit for braking on a dry road according to \cite{kesting2007general}). In this study, we consider acceleration as a real valued quantity that may also be negative (i.e., braking).

Our model adopts a simplified view of a lane and vehicle occupancy, and how accidents are detected, omitting complexities such as vehicle and lane widths. Incorporating these factors would only add unnecessary complication to the model without significantly affecting the observed behavior. In our model, a vehicle either completely occupies a lane or is rapidly transitioning between two lanes. Thus, an accident is recognized when two vehicles either share the same lane or are transitioning to or from identical lanes and intersect longitudinally.

\section{Demonstrative study}
\begin{figure*}
    \centering
    \begin{subfigure}[b]{0.75\textwidth}
        \centering
        \includegraphics[width=\textwidth]{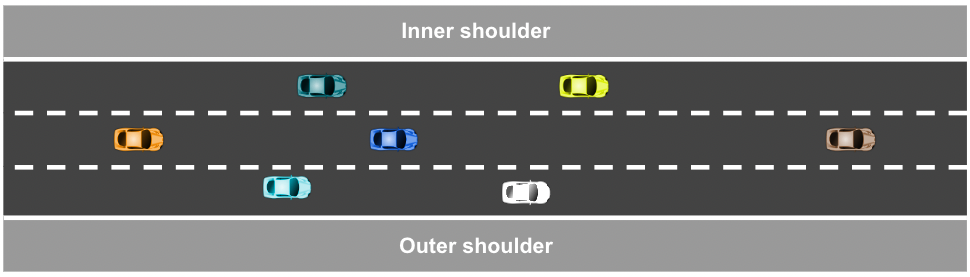}
        \caption[Network2]%
        {{\small An example freeway with both inner and outer shoulders present}}    
        \label{fig:bothShoulder}
    \end{subfigure}
    \vskip\baselineskip
    \begin{subfigure}[b]{0.75\textwidth}
        \centering
        \includegraphics[width=\textwidth]{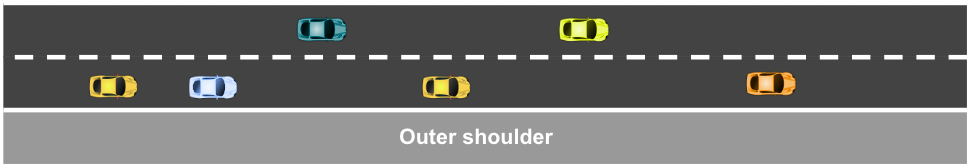}
        \caption[Network2]%
        {{\small An example freeway with only an outer shoulder and no inner shoulder}}    
        \label{fig:oneShoulder}
    \end{subfigure}
    \vskip\baselineskip
    \begin{subfigure}[b]{0.75\textwidth}   
        \centering 
        \includegraphics[width=\textwidth]{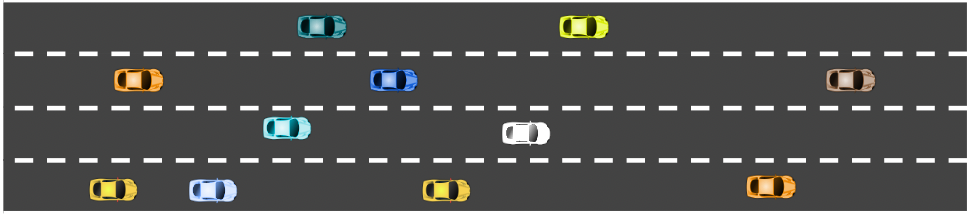}
        \caption[]%
        {{\small An example freeway with no shoulders}}
        \label{fig:noShoulder}
    \end{subfigure}

    \caption[ The average and standard deviation of critical parameters ]
    {\small We examine the relative accident rates for all combinations of the number of lanes and the existence or lack of inner and/or outer shoulders. } 
    \label{fig:laneCombinations}
\end{figure*}

In our demonstrative study, the goal is to observe whether increasing the number of lanes from two up to five will have a positive or negative impact on the accident rates for the three named accident types. The accident rates we will observe will be normalized with the 2-lane freeways accident rates. We consider all possible combinations of 2-lane up to 5-lane freeways and possible inner and/or outer shoulders. Fig. \ref{fig:laneCombinations} depicts some example combinations. 
The model in the previous section can be used to examine various accident scenarios. In this part, we formulate and present the accident types we used for demonstrating the use of the model as well as the particulars of our simulation. 

\subsection{Accident types investigated}
We investigate three prevalent accident types and see how increasing the number of lanes will increase or decrease the rate of accidents relative to the baseline accident rate for 2 lanes. 

\subsubsection{Following too closely}
We model this accident type by having a trigger car suddenly brake with maximum deceleration $B_\text{critical}$ at time $t=0$. The following car will automatically try to avoid an accident by braking or changing lanes. For this accident type, we set $R=0.7$. We record the rate of accidents (Fig. \ref{fig:accidentTypeFollowTooClosely}). 

\subsubsection{Driver inattention}
Similarly, we model this accident type by having the trigger car above braking suddenly at $t=0$. Furthermore, if a follower car exists in the same lane (i.e. the inattentive driver) , it will continue with its initial speed and only becomes aware of the traffic condition after a delay of $T_d=0.5s$. That is, the delayed reaction is a proxy for the inattention. For this accident type, we set $R=0.2$. The rate of accidents is recorded (Fig. \ref{fig:accidentTypeDriverInattention}).

\subsubsection{Unsafe change of lanes}
\begin{figure*}
    \centering
    \begin{subfigure}[b]{0.3\textwidth}
        \centering
        \includegraphics[width=\textwidth]{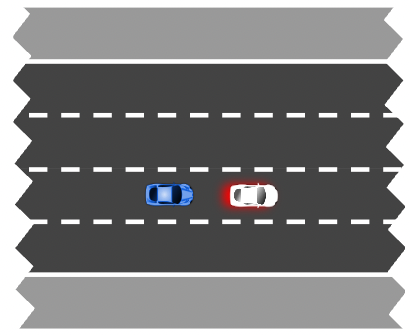}
        \caption[Network2]%
        {{\small In the "following too closely" accident scenario, a trigger vehicle will brake suddenly, creating a potentially dangerous situation if another vehicle is following too closely.}}    
        \label{fig:accidentTypeFollowTooClosely}
    \end{subfigure}
    \hfill
    \begin{subfigure}[b]{0.3\textwidth}  
        \centering 
        \includegraphics[width=\textwidth]{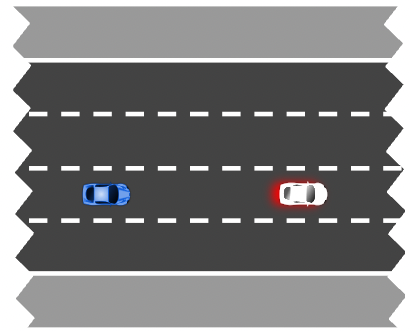}
        \caption[]%
        {{\small In the "driver inattention" accident scenario, a trigger vehicle will brake suddenly, but the follower car, unaware of the change, will react with a delay that might cause an accident. }}    
        \label{fig:accidentTypeDriverInattention}
    \end{subfigure}
    \hfill
    \begin{subfigure}[b]{0.3\textwidth}  
        \centering 
        \includegraphics[width=\textwidth]{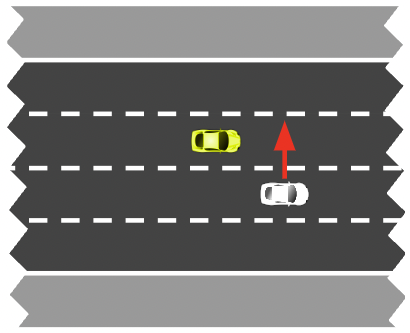}
        \caption[]%
        {{\small In the "unsafe change of lanes" accident scenario, a trigger vehicle will change lanes abruptly, without concern for the amount of deceleration the second vehicle will need to avoid an accident.}}    
        \label{fig:accidentTypeUnsafeChangeOfLanes}
    \end{subfigure}
    \caption[ The average and standard deviation of critical parameters ]
    {\small We estimate the relative rate of accidents for three accident types of interest in this paper as depicted here.} 
    \label{fig:accidentTypes}
\end{figure*}

We model this accident type as follows. For this type of accident as well, we have a (dummy) trigger car similar to the above, braking at time $t=0$. However, to introduce possible speed variation (since every car is driving at the same speed), if a follower car exists (the actual trigger car), it will wait for at least $T_d= 0.5s$ and after that on every simulation time step ($dt=0.05s$), it will attempt to switch lanes from the original lane with probability $0.1s/dt$ until the lane is switched. 
When switching lanes, the vehicle will not respect the safety rule of Eq. \ref{eqSafeLaneChangeState} and will only avoid a lane change if it results in an immediate accident (such as when another vehicle is driving almost in parallel in the target lane). We set $R=0.2$ as well for this accident type. It is worth noting a few details: If by the end, the following car cannot switch lanes and crashes into the trigger car, we will disqualify the sample. Furthermore, once the abrupt lane change happens, the dummy trigger car is removed (Fig. \ref{fig:accidentTypeUnsafeChangeOfLanes}).

\subsection{Simulation setup and objectives}
In our setup, we randomly place cars in a small window ($75m$) of a one-way freeway for a short time ($20s$). We place a car that will trigger an accident at the $15m$ mark from the front of the window. We allow a long enough empty runway for the dynamics to unfold ($685m$ in this case). Each vehicle has a length of $4m$ and is initially separated by at least $2m$ from other vehicles. Every vehicle will start at the same speed as follows. We update the simulation variables with a time step of $dt=0.05s$. We have provided more details regarding the random placement of cars in the appendix.

In our simulation, we consider the peak density ($25.37$ veh/km, speed: $53.5\text{kmh}^{-1}$) and speed from US-290E at Woodbury Drive in Austin, Texas according to the research done by Azimi et al. in \cite{azimi2010categorizing} on real traffic data. 

In terms of the number of shoulders, we consider three different types of one-way freeways in our simulation: 1. freeways with no shoulders, 2. freeways with one shoulder, 3. freeways with two shoulders (Fig. \ref{fig:laneCombinations}).

We will perform a Monte Carlo simulation and record the frequency of accidents for one-way freeways with 2, 3, 4, and 5 lanes. We will take the accident rate in freeways with 2 lanes as our baseline and calculate the relative increase or decrease in accident frequency as we add more lanes. In doing so, we consider every possible combination of shoulders type and the number of lanes.

\section{Results and discussion}

\begin{figure*}
    \centering
    \begin{subfigure}[b]{0.6\textwidth}
        \centering
        \includegraphics[trim=0 520 0 0,clip,width=\textwidth]{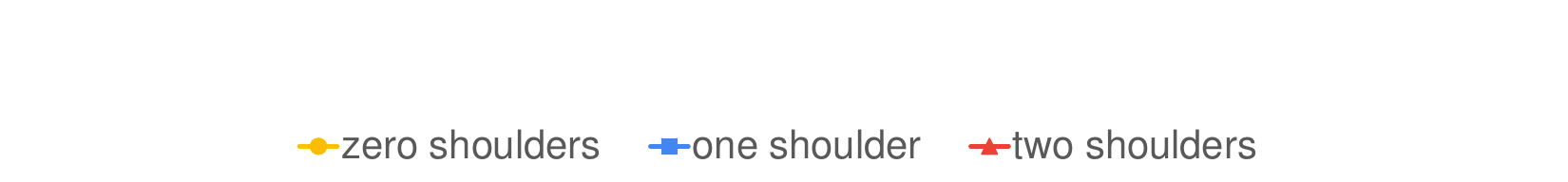}
    \end{subfigure}
    \centering
    \begin{subfigure}[b]{0.6\textwidth}
        \centering
        \includegraphics[trim=0 50 0 60,clip,width=\textwidth]{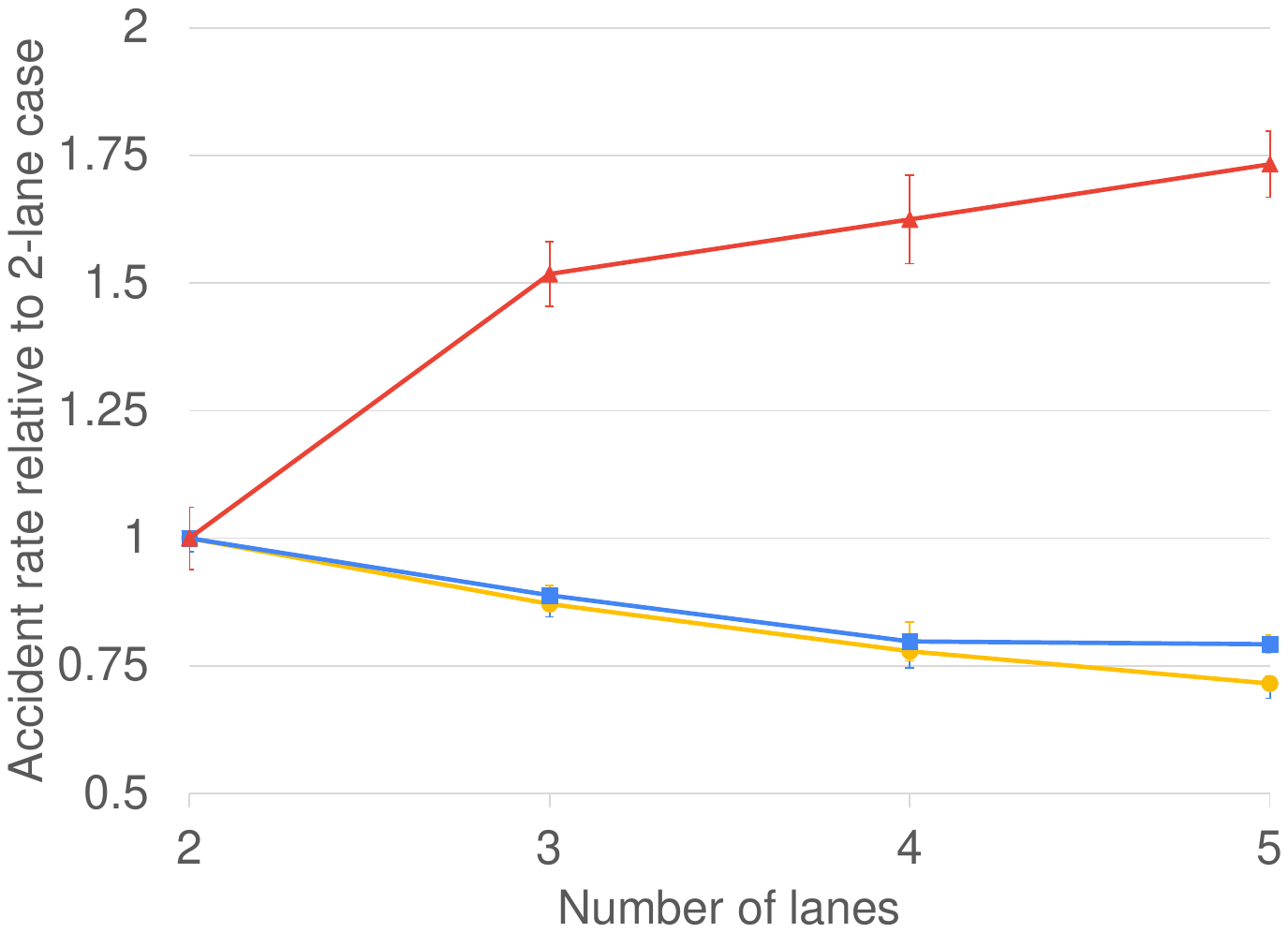}
        \caption[Network2]%
        {{\small The accident type "following too closely"}}    
        \label{fig:followTooCloseResults-high}
    \end{subfigure}
    \vskip\baselineskip
    \begin{subfigure}[b]{0.6\textwidth}
        \centering
        \includegraphics[trim=0 50 0 60,clip,width=\textwidth]{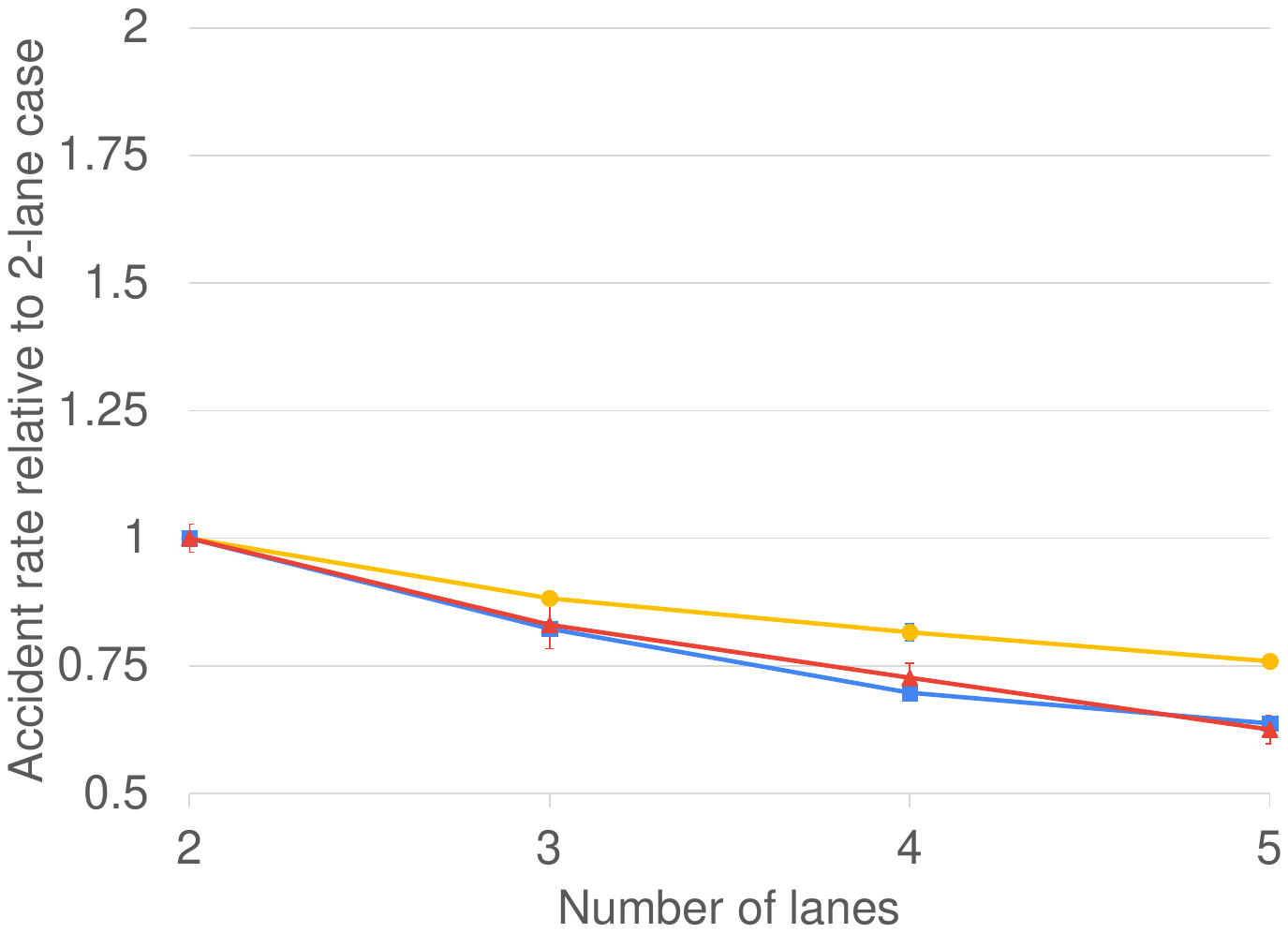}
        \caption[Network2]%
        {{\small  The accident type "driver inattention"}}    
        \label{fig:inattentionResults-high}
    \end{subfigure}
    \vskip\baselineskip
    \begin{subfigure}[b]{0.6\textwidth}
        \centering
        \includegraphics[trim=0 50 0 60,clip,width=\textwidth]{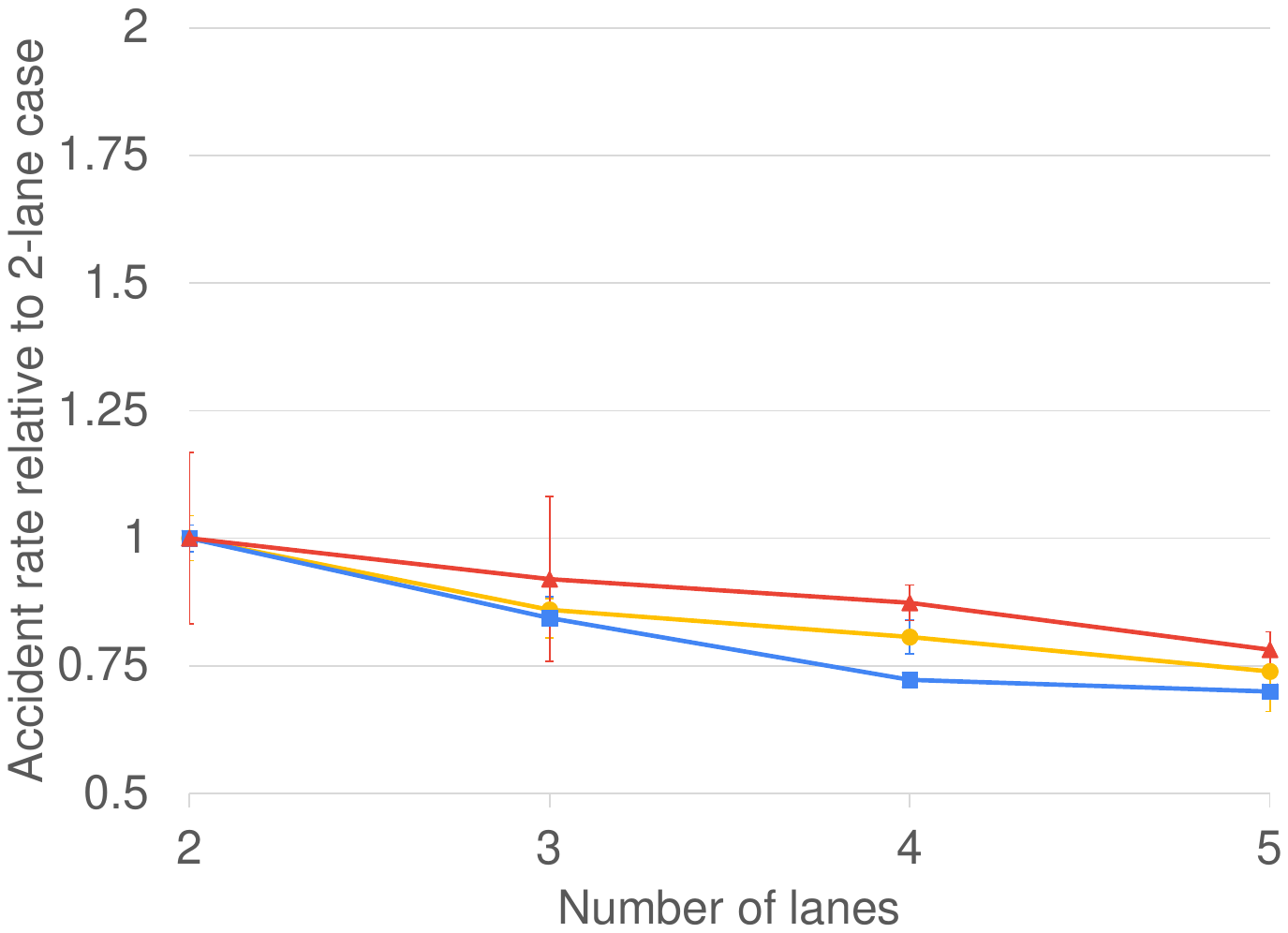}
        \caption[Network2]%
        {{\small  The accident type "unsafe change of lanes"}}    
        \label{fig:unsafeResults-high}
    \end{subfigure}

    \caption[ The average and standard deviation of critical parameters ]
    {\small In each figure, for various accident types, the accident rate relative to the 2-lane accident rate is shown for freeways with zero, one, or two shoulders. } 
    \label{fig:allResults}
\end{figure*}

 Fig. \ref{fig:allResults} shows the results of simulations for the three accident types from the previous section as we increase the number of lanes. In each graph, the case of zero, one, and two shoulders are represented.

 In the case of "following too closely", we see two different behaviors when two shoulders are present vs. when there are one or none. In the case of one or no shoulders, we see a general decline in the relative "following too closely" accident rates. This is possibly due to a higher degree of freedom for evasive maneuvers. The sharpest drop occurs after adding one lane to a two-lane highway with generally slower drop with more lanes. In the case with two shoulders, however, we see an increase in the accident rates as the number of lanes increases. A possible explanation for this effect can be that adding more lanes, will result in the shoulders being further away from the middle lanes and hence making them harder to reach for the vehicles on those lanes when needed.

Similarly, in the case of "driver inattention", it can be seen that in all cases, adding more lanes will result in fewer accidents. Going from 2 lanes to 3 lanes has the largest impact while subsequent addition of lanes results in a more modest drop in the accident rates for the cases investigated.

Finally, in the case of "unsafe change of lanes", adding more lanes will result in fewer accidents. The largest drop in accident rates happen with adding the 3rd lane. 

\section{Limitations}

Similar to \cite{hou2020study}, a
limitation related to our work is that our microscopic model is not calibrated and validated with real-world data. There is a lack of uniform data, related to before / after studies of the same freeways upon expansion to more lanes. Nevertheless, if such data becomes more widely available in the future, the time-varying parameters (such as the demand, etc.) of the freeway will not remain the same. If this data becomes available, it might be possible to control for the time-varying external factors to some extent and use the data to calibrate and validate the model given an enormous of computational power is available. Furthermore, accidents are very rare and they usually stem from a chain of unfortunate events. Therefore, to calibrate the model, even if all the needed data is present will require a massive amount of computation which can be a limiting factor for most studies in this area and ours is no exception.

\section{Conclusions}
The core question we strove to answer in this research was whether a freeway with more lanes is safer than one with fewer lanes. We devised a microscopic model that is capable of causing accidents to study the safety effect of adding lanes. This allowed us to model the fundamentals of three different common accident types on urban freeways, such as "following too closely", "driver inattention", and "unsafe change of lanes". We observed that whether shoulders are present or not can even change the positive effect of adding lanes to negative in some cases. Furthermore, to reduce the computational burden of collecting many samples, we simulated each possible incident with only a few cars in a short segment of the freeway and for a few seconds from the time a triggering event started. This allowed us to collect a substantial number of samples which would have been otherwise not easily possible.

\section*{Declaration of Competing Interest}
\textbf{Mirmojtaba Gharibi} - no conflicts of interest.

\textbf{John-Paul Clarke} - no conflicts of interest.


\appendix
\section{Appendix}
\subsection{Initial random placement of cars}
For the initial random placement of the cars on the road, we multiply the density by the number of lanes and round the result to give the total number of cars to put on the freeway of length 1km. Each car's initial starting position is at a whole number as a simplifying approximation. In each iteration, a car is placed at a uniformly random index on the freeway represented by a lattice (just for the placement; for the simulation we use continuous simulation), only if the 2m minimum distance between the placed car and existing cars on the lattice is satisfied. We then remove all the cars that are outside the 75m window of interest. For a large enough window and as the initial length of the highway goes to infinity, the number of cars in the considered window can be approximated with a Poisson distribution.


\bibliographystyle{cas-model2-names}

\bibliography{cas-refs}

\newpage

\bio{figs/MGharibi}
\noindent\textbf{\underline{Mirmojtaba Gharibi}}, received the B.ASc. in electrical engineering from Sharif University of Technology in 2009. He completed his M.Math, and PhD degree both in Computer Science at University of Waterloo, Waterloo, Canada. He is currently affiliated with the University of Texas at Austin.
\endbio

\section*{}

\bio{figs/JPClarke}
\noindent\textbf{\underline{John-Paul Clarke}} is a professor of Aerospace Engineering and Engineering Mechanics at The University of Texas at Austin, where he holds the Ernest Cockrell Jr. Memorial Chair in Engineering. Before joining the faculty at UT Austin, he was a faculty member at Georgia Tech, the Vice President of Strategic Technologies at United Technologies Corporation (now Raytheon), a faculty member at MIT, and a researcher at Boeing and NASA JPL. He has also co-founded multiple companies, most recently Universal Hydrogen – a company dedicated to the development of a comprehensive carbon-free solution for aviation.
\endbio

\end{document}